\documentclass[12pt,english]{article}
\pdfoutput=1
\usepackage{lmodern}

\usepackage[T1]{fontenc}
\usepackage[latin9]{inputenc}
\usepackage{geometry}
\usepackage{color}
\usepackage{babel}
\usepackage{cite}
\usepackage{mathtools}
\usepackage{amsmath}
\usepackage{amssymb}
\usepackage{graphicx}
\usepackage{esint}
\usepackage{bbm}
\usepackage[hang,flushmargin]{footmisc} 
\usepackage[unicode=true,pdfusetitle,
 bookmarks=true,bookmarksnumbered=false,bookmarksopen=false,
 breaklinks=false,pdfborder={0 0 1},backref=false,colorlinks=true]
 {hyperref}
 \hypersetup{pdftitle={Proof of the Weak Gravity Conjecture from Black Hole Entropy}, pdfauthor={Clifford Cheung, Junyu Liu, Grant N. Remmen}, citecolor=black,linkcolor=black,urlcolor=black}
\makeatletter

 \@ifundefined{textcolor}{}
 {
   \definecolor{BLACK}{gray}{0}
   \definecolor{WHITE}{gray}{1}
   \definecolor{RED}{rgb}{1,0,0}
   \definecolor{GREEN}{rgb}{0,1,0}
   \definecolor{BLUE}{rgb}{0,0,1}
   \definecolor{CYAN}{cmyk}{1,0,0,0}
   \definecolor{MAGENTA}{cmyk}{0,1,0,0}
   \definecolor{YELLOW}{cmyk}{0,0,1,0}
 }
\usepackage{babel}
\makeatletter
\def\simgt{\mathrel{\lower2.5pt\vbox{\lineskip=0pt\baselineskip=0pt
           \hbox{$>$}\hbox{$\sim$}}}}
\def\simlt{\mathrel{\lower2.5pt\vbox{\lineskip=0pt\baselineskip=0pt
           \hbox{$<$}\hbox{$\sim$}}}}
\makeatother

\newcommand{\be}{\begin{equation}}
\newcommand{\ee}{\end{equation}}
\newcommand{\bea}{\begin{eqnarray}}
\newcommand{\eea}{\end{eqnarray}}
\newcommand{\eq}[2]{\be\begin{aligned}#1 \label{#2}\end{aligned}\ee}

\newcommand{\Ref}[1]{Ref.~\cite{#1}}
\newcommand{\Fig}[1]{Fig.~\ref{#1}}
\newcommand{\Eq}[1]{Eq.~\eqref{#1}}
\newcommand{\Eqs}[2]{Eqs.~\eqref{#1} and \eqref{#2}}
\newcommand{\Sec}[1]{Sec.~\ref{#1}}

\newcommand{\App}[1]{App.~\ref{#1}}

\newcommand{\LL}{\mathcal{L}}

\newcommand{\BHentropy}{\frac{2\pi  \wt A}{\kappa^2}}
\newcommand{\wt}[1]{\widetilde{#1}}

\renewcommand{\d}{c}
\newcommand{\e}{d}

\begin{document}

\interfootnotelinepenalty=10000
\baselineskip=18pt

\hfill CALT-TH-2018-007
\hfill

\vspace{2cm}
\thispagestyle{empty}
\begin{center}
{\LARGE \bf
Proof of the Weak Gravity Conjecture \\  \medskip from Black Hole Entropy
}\\
\bigskip\vspace{1cm}{
{\large Clifford Cheung,${}^a$ Junyu Liu,${}^a$ and Grant N. Remmen${}^b$}
} \\[7mm]
 {\it ${}^a$Walter Burke Institute for Theoretical Physics \\
    California Institute of Technology, Pasadena, CA 91125 \\[1.5mm]
${}^b$Center for Theoretical Physics and Department of Physics\\
    University of California, Berkeley, CA 94720 and \\
Lawrence Berkeley National Laboratory, Berkeley, CA 94720} \let\thefootnote\relax\footnote{email: \url{clifford.cheung@caltech.edu}, \url{jliu2@caltech.edu}, \url{grant.remmen@berkeley.edu}} \\
 \end{center}
\bigskip
\centerline{\large\bf Abstract}

\begin{quote} \small

We prove that higher-dimension operators contribute positively to the entropy of a thermodynamically stable black hole at fixed mass and charge.  Our results apply whenever the dominant corrections originate at tree level from quantum field theoretic dynamics.
More generally, positivity of the entropy shift is equivalent to a certain inequality relating the free energies of black holes.  
These entropy inequalities mandate new positivity bounds on the coefficients of higher-dimension operators.  One of these conditions implies that the charge-to-mass ratio of an extremal black hole asymptotes to unity from above for increasing mass.  Consequently, large extremal black holes are unstable to decay to smaller extremal black holes and
the weak gravity conjecture is automatically satisfied.  Our findings generalize to arbitrary spacetime dimension and to the case of multiple gauge fields.  The assumptions of this proof are valid across a range of scenarios, including string theory constructions with a dilaton stabilized below the string scale.
\end{quote} 

\begin{quote} \small

\end{quote}
\newpage

\tableofcontents

\newpage

\setcounter{footnote}{0}

\section{Introduction}\label{sec:intro}

In this paper we argue that black hole thermodynamics implies new constraints on the coefficients of higher-dimension operators.  Our results are based on a certain universal property of entropy.  In particular, consider a system ${\cal T}$ prepared in a macrostate whose microstate degeneracy is quantified by entropy $S$.  Now let us define the system $\wt{\cal T}$ to be a restriction of ${\cal T}$ in which a handful of degrees of freedom have been frozen to a fixed configuration.  If $\wt{\cal T}$ is prepared in the same macrostate as ${\cal T}$, then the corresponding entropy $\wt S$ will be less than $S$ because the microstate degeneracy is diminished.  Thus, we find that
\eq{
S = \wt S + \Delta S ,
}{eq:S_def}
where the entropy shift $\Delta S$ is strictly positive.    

A similar logic applies to  black hole entropy.  Consider the low-energy effective field theory of photons and gravitons in general spacetime dimension $D$, defined at a scale far below the mass gap.  The effective Lagrangian is
\eq{
{\cal L} =\wt {\cal L} +\Delta {\cal L},
}{eq:Ltotal}
where the first term describes pure Einstein-Maxwell theory,\footnote{Throughout, we work in units where $\kappa^2 = 8\pi G$, metric signature $(-,+,\cdots,+)$, and sign conventions $R_{\mu\nu} = R^{\rho}_{\;\;\mu\rho\nu}$ and $R^{\mu}_{\;\;\nu\rho\sigma} = \partial_\rho \Gamma^\mu_{\nu\sigma} -\partial_\sigma \Gamma^\mu_{\nu\rho} + \Gamma^\mu_{\rho\alpha}\Gamma^\alpha_{\nu\sigma} - \Gamma^\mu_{\sigma\alpha}\Gamma^\alpha_{\nu\rho}$.}
 \eq{
\wt{\cal L} = \frac{1}{2\kappa^2} R - \frac{1}{4}F_{\mu\nu}F^{\mu\nu},
}{eq:EM}
and second term encodes corrections from higher-dimension operators,
\eq{
\Delta {\cal L} = &
\phantom{{} + {}} 
\d_1 R^2 +\d_2 R_{\mu\nu}R^{\mu\nu} + \d_3 R_{\mu\nu\rho\sigma}R^{\mu\nu\rho\sigma} \\
& + \d_4 RF_{\mu\nu}F^{\mu\nu} + \d_5 R_{\mu\nu}F^{\mu\rho}F^\nu_{\;\;\rho} + \d_6 R_{\mu\nu\rho\sigma}F^{\mu\nu}F^{\rho\sigma} \\
& + \d_7 F_{\mu\nu}F^{\mu\nu}F_{\rho\sigma}F^{\rho\sigma} + \d_8 F_{\mu\nu}F^{\nu\rho}F_{\rho\sigma}F^{\sigma\mu}.
}{eq:L}
Without loss of generality, we have dropped all terms involving $\nabla_\rho F_{\mu\nu}$, which are equivalent via the Bianchi identities to terms already accounted for or terms involving $\nabla_\mu F^{\mu\nu}$  \cite{Deser:1974cz}, which vanish in the absence of charged matter sources, which we assume throughout.  

For our analysis, we study a large black hole of fixed mass $m$ and charge $q$ as measured in natural units at spatial infinity.  In the presence of higher-dimension operators, the metric is
\eq{
g_{\mu\nu} = \wt g_{\mu\nu} + \Delta g_{\mu\nu}.
}{}
Unless otherwise stated, all variables with tildes like $\wt g_{\mu\nu}$ will denote quantities corresponding to a Reissner-Nordstr\"om black hole of the same mass and charge in pure Einstein-Maxwell theory, while variables with deltas like $\Delta g_{\mu\nu}$ will denote the leading corrections from higher-dimension operators, which are linear in the coefficients $\d_i$.

It is straightforward to  compute the black hole entropy using the Wald formula \cite{Wald:1993nt},\footnote{A general formula for entanglement entropy has also been proposed \cite{Dong:2013qoa}, but this reduces to the Wald formula for the static Killing horizons relevant to our analysis.}
\eq{
 S=-2\pi \int_{\Sigma} \frac{\delta {\mathcal{L}}}{\delta {{R}_{\mu\nu\rho\sigma}}}  \, {{\epsilon }_{\mu\nu}}{{\epsilon }_{\rho\sigma}} ,
}{eq:Waldgeneral}
where the integration region $\Sigma$ is the horizon and $\epsilon_{\mu\nu}$ is the binormal to the horizon, normalized so that $\epsilon_{\mu\nu}\epsilon^{\mu\nu}=-2$.       A large portion of our technical analysis will be the explicit evaluation of \Eq{eq:Waldgeneral} at fixed $q$ and $m$.  The entropy shift $\Delta S$ is then defined according to \Eq{eq:S_def}, where
\eq{
\wt S = \frac{\wt A}{4G} = \BHentropy 
}{eq:BHentropy}
is the Bekenstein-Hawking entropy  \cite{Bekenstein:1973ur,Hawking:1976de}.  Deviations from \Eq{eq:BHentropy} arise from higher-dimension operator corrections to the Lagrangian, ${\cal L} = \wt{\cal L} + \Delta {\cal L}$, and the area of horizon, $A = \wt A + \Delta A$.

Using standard thermodynamic identities we show that corrections to the Bekenstein-Hawking entropy of a black hole at fixed mass and charge satisfy 
\eq{
\Delta S > 0
}{eq:inequality}
whenever the free energy of the black hole is less than that of a Reissner-Nordstr\"om black hole at the same temperature,
\eq{
F(\beta) < \wt F(\beta).
}{eq:F_inequality}
Using Euclidean path integral methods we then prove \Eq{eq:F_inequality}
for {\it i}) a thermodynamically stable black hole  in {\it ii}) a theory in which the dominant contributions to higher-dimension operators are generated at tree level by massive quantum fields.
The underlying logic of our proof mirrors the parable of ${\cal T}$ and $\wt {\cal T}$ discussed previously.  In particular,
$\wt {\cal L}$ is obtained directly from the ultraviolet completion of ${\cal L}$ by placing a restriction on massive field fluctuations in the Euclidean path integral.  The difference between $S$ and $\wt S$ then quantifies the entropy contributions from these modes.

Condition {\it i}) holds for positive specific heat \cite{Davies:1978mf},  excluding from consideration the Schwarzschild black hole but permitting Reissner-Nordstr\"om black holes over a range of charge-to-mass ratios, $q/m > \sqrt{2D-5}/(D-2)$.  Condition {\it ii}) arises naturally in a number of physically-motivated contexts such as string theory, which typically predicts the existence of gravitationally-coupled scalars like the dilaton.   If supersymmetry breaking occurs below the string scale, then these states are well described by quantum field theory and can be integrated out at tree level to generate the higher-dimension operators in \Eq{eq:L}. Note that condition {\it ii}) is perfectly consistent with additional corrections entering at loop level or from intrinsically stringy dynamics, provided these contributions are parametrically smaller than the tree-level component.

While we have only proven the free energy condition in \Eq{eq:F_inequality} under certain assumptions, its connection to positivity of the entropy shift in \Eq{eq:inequality} is robust and completely general.     Furthermore, a trivial corollary to \Eq{eq:inequality} is a positivity condition on the differential entropy generated at each mass threshold encountered while flowing to the infrared.

Remarkably, classical entropy corrections dominate over quantum contributions over a broad range of black hole masses still within the regime of validity of the effective field theory.  Within this window, our proof of \Eq{eq:inequality} applies.  Since the shift in entropy depends on the coefficients of higher-dimension operators, we derive a new class of positivity bounds on the effective field theory.
This produces a one-parameter family of constraints on the corresponding coefficients $\d_i$ labeled by the charge-to-mass ratio  of the black hole from which the bound was derived. Although these conditions are derived from a very particular black hole construction, the resulting positivity bounds constrain the coefficients $\d_i$ in general and are independent of the background.

\begin{figure}[t]
\begin{center}
\hspace{-10mm} \includegraphics[height=0.6\textwidth]{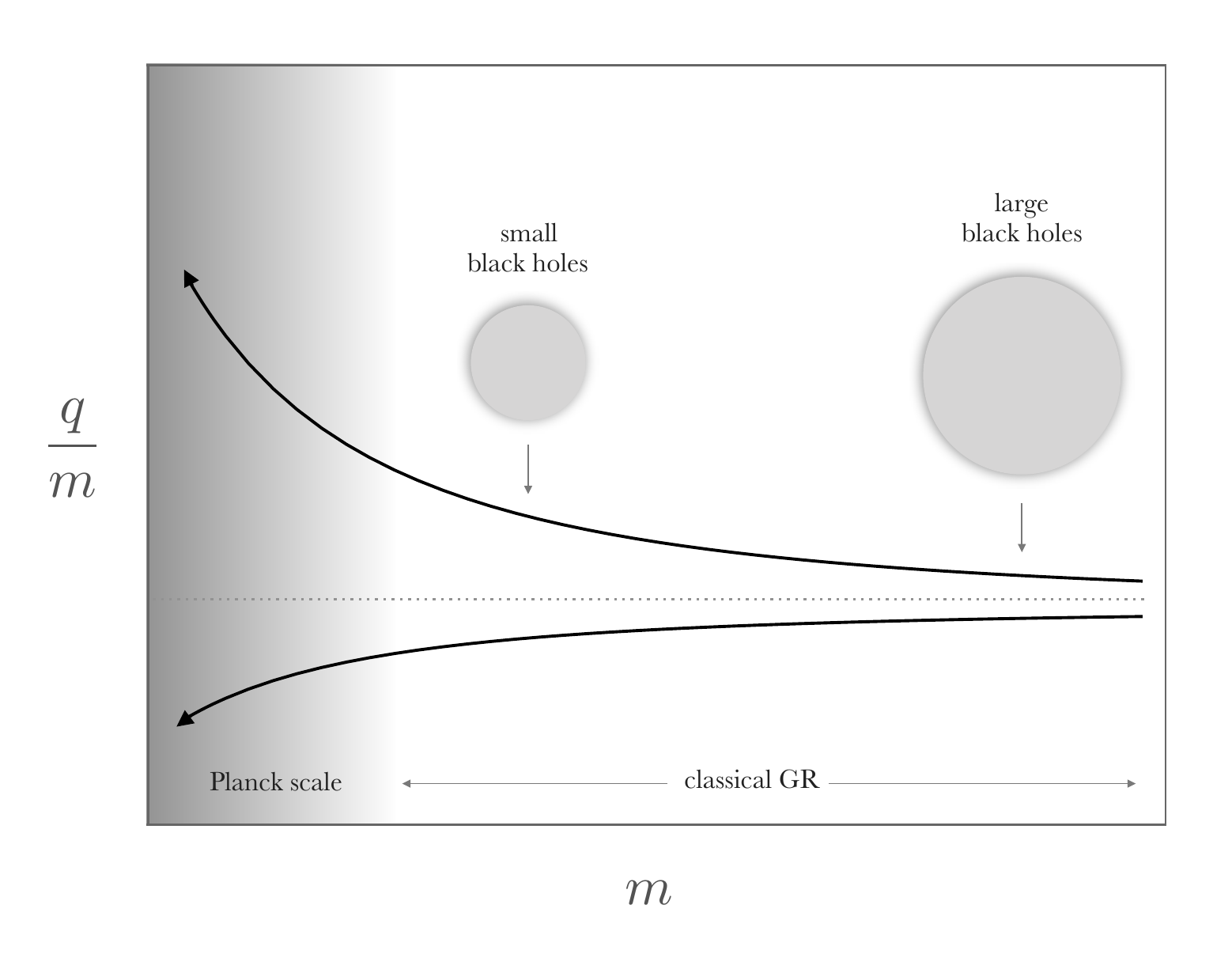}
\end{center}
\vspace{-10mm}
\caption{Black holes of maximal charge shown as a function of mass $m$ and charge-to-mass ratio $q/m$.    Higher-dimension operators induce corrections to the extremality condition. If these corrections are positive, then the WGC is automatically satisfied (upper solid curve) since large black holes are unstable to decay to smaller ones.   If these corrections are negative (lower solid curve), then the WGC mandates additional light, superextremal particles to avoid an infinite number of stable extremal black hole remnants.}
\label{fig:mass_plot}
\end{figure}

For the case of highly charged black holes, we obtain a positivity condition on a very specific combination of higher-dimension operator coefficients.  Remarkably, it is this exact combination of parameters that also enters into the higher-dimension operator correction to the extremality condition for black holes.  In particular, we find that the charge-to-mass ratio for an exactly extremal black hole satisfies
\eq{
\frac{q}{m} - 1 \propto \Delta S,
}{eq:qm1}
where the right-hand side is positive by \Eq{eq:inequality}.  Since higher-dimension operators decouple at long distances, the charge-to-mass ratio asymptotes to unity {\it from above} as we consider larger and larger extremal black holes.  Thus, from charge and energy conservation, it is always possible for an extremal black hole to decay to smaller extremal black holes of marginally higher charge-to-mass ratio, as shown in \Fig{fig:mass_plot}.  Notably, the existence of such states is precisely mandated by the weak gravity conjecture (WGC), which asserts that an Abelian gauge theory consistently coupled to gravity must contain a state whose charge exceeds its mass in Planck units \cite{ArkaniHamed:2006dz}, so
\eq{
\frac{q}{m}>1.
}{eq:WGCclaim}
The motivation for the WGC is to forbid the existence of an infinite number of stable states not protected by symmetry.  
The main result of this paper is that this bound is automatically satisfied by small black holes.  Though mysterious at first glance, the connection between entropy and extremality in \Eq{eq:qm1} actually follows immediately from the near-horizon properties of the metric for a charged black hole.  As we will show, this connection enables us to straightforwardly extend all of our arguments to the multi-charge generalization of the WGC \cite{Cheung:2014vva} in arbitrary dimension $D$.

The WGC is satisfied in numerous concrete examples and is strongly motivated by folk theorems forbidding exact global symmetries that arise in the vanishing-charge limit. Moreover, the WGC is a celebrated avatar of the so-called swampland program \cite{Vafa:2005ui,Ooguri:2006in,ArkaniHamed:2006dz,Adams:2006sv}, whose ultimate aim is to systematically delineate regions in effective field theory space consistent with quantum gravitational ultraviolet completion.  

Strictly speaking, a bona fide swampland condition distinguishes between low-energy effective field theories that from an infrared standpoint are otherwise entirely consistent.  From this perspective it is unclear whether a theory that fails the WGC is in the swampland or {\it merely pathological} in a sense that can be diagnosed purely from low-energy considerations.
For this reason, a related effort has sought to exclude regions of the low-energy parameter space using infrared consistency, e.g., constraints from causality, unitarity, and analyticity of scattering amplitudes \cite{Adams:2006sv,Cheung:2014ega,Bellazzini:2015cra,Cheung:2016yqr,GB+}.    While the WGC has previously eluded a formal general proof, the present work demonstrates that it is mathematically equivalent to a certain well-motivated---and in many circumstances provable---property of black hole entropy. 

The remainder of this paper is organized as follows.  In \Sec{sec:proof} we prove, given certain assumptions, that corrections to the Bekenstein-Hawking entropy are positive.  Afterwards, in  \Sec{sec:UV} we discuss how various contributions to the black hole entropy arise and estimate their relative size.  In order to constrain the coefficients of higher-dimension operators, we restrict to black holes within a certain mass range.  Next, in \Sec{sec:spacetime} we present the perturbative solution for a charged black hole in the presence of higher-dimension operators.  We then compute the black hole entropy in \Sec{sec:entropy} and translate the positivity condition on entropy into a new class of bounds on higher-dimension operator coefficients in \Sec{sec:bound}.  After discussing the implications of these results for the WGC, we conclude in \Sec{sec:conclusions}.

\section{Proof of  $\Delta S >0$} \label{sec:proof}

In this section, we study corrections to the Bekenstein-Hawking entropy of a thermodynamically stable black hole at fixed mass and charge.  We prove that these contributions are positive whenever they come from higher-dimension operators generated at tree level by quantum fields.

\subsection{Assumptions} \label{sec:assumptions}

Let us state our assumptions explicitly.  First, we assume the existence of quantum fields $\phi$ at a characteristic mass scale $m_\phi \ll \Lambda$, where $\Lambda$ is the energy scale at which quantum field theory breaks down.  This parametric separation is required so that quantum field theory has some regime of validity.  By the usual rules of effective field theory, the higher-dimension operators in \Eq{eq:L} receive small contributions suppressed by the cutoff $\Lambda$.  In general, $\Lambda$ can be parametrically smaller than the Planck scale.

Second, we assume that the $\phi$ degrees of freedom couple to photons and gravitons such that integrating them out generates the higher-dimension operators in \Eq{eq:L} classically, i.e., at tree level.  Since $m_{\phi} \ll \Lambda$, these states induce the dominant contributions to the higher-dimension operators in \Eq{eq:L}.   Specifically, the corresponding operator coefficients scale as  $\d_i \propto 1/m_\phi^2 \gg 1/\Lambda^2$ since tree-level $\phi$ exchange is always accompanied by a single factor of $1/m_\phi^2$ coming from the propagator denominator.   Thus, effects arising from the cutoff $\Lambda$ will be negligible in any context in which quantum field theory is applicable.

As noted previously, states like $\phi$ are a common feature in string theory, whose low energy spectrum includes particles like the dilaton and moduli, which are massless in the supersymmetric limit. In the presence of supersymmetry breaking, these flat directions are lifted, thus inducing masses $m_\phi \ll \Lambda$, where $\Lambda$ is the string scale.

While our arguments are perfectly consistent with a scale $\Lambda$ far below the Planck scale, we will frequently refer to pure Einstein-Maxwell theory and the pure Reissner-Nordstr\"om solution as a baseline of comparison.  We do so entirely out of convenience and not because any component of our argument requires that quantum field theory be applicable up to the Planck scale.  Hence, in an abuse of nomenclature, we hereafter refer to the higher-dimension operator contributions of order $1/m_\phi^2$ as corrections to pure Einstein-Maxwell theory, bearing in mind that we actually mean pure Einstein-Maxwell theory {\it plus} contributions of order $1/\Lambda^2$, which are parametrically smaller than all the contributions of interest. 

Third, we focus on black holes that are thermodynamically stable, i.e., have positive specific heat.  As we will see, this is necessary for technical reasons so that we can exploit certain properties of the Euclidean path integral.

\subsection{Positivity Argument}

Consider a positively charged black hole of mass $M$ and charge $Q$ perturbed by higher-dimension operator corrections in general spacetime dimension $D$.  
As we will show in detail in \Sec{sec:spacetime}, the perturbed metric  $g_{\mu\nu}  = \wt g_{\mu\nu} +\Delta g_{\mu\nu}$ can be computed from the perturbed Lagrangian ${\cal L} =\wt {\cal L} +\Delta {\cal L}$, where unless otherwise stated all quantities are expressed as perturbations on a Reissner-Nordstr\"om black hole of the same mass $M$ and charge $Q$ in pure Einstein-Maxwell theory.

 From the perturbed entropy we can define the corresponding inverse temperature $\beta = \partial_M S $, which we write as
\eq{
 \beta = \wt{\beta} + \Delta \beta ,
}{eq:beta_def}
where $\wt \beta = \partial_M \wt S$ is the inverse temperature of the Reissner-Nordstr\"om black hole and $\Delta \beta = \partial_M \Delta S$ is the shift of the inverse temperature.  The latter appears because higher-dimension operators will in general change the temperature of a black hole at a fixed mass and charge.

Next, we compute the free energy $F(\beta)$ of the perturbed black hole in a canonical ensemble at inverse temperature $\beta$.  The free energy is calculated from the Euclidean path integral,
\eq{
e^{-\beta F(\beta)} =  Z(\beta) = \int d[\hat g]d[ \hat A] \, e^{-I[\hat g, \hat A]},
}{}
where $\hat g$ and $\hat A$ are integration variables running over metric and gauge field configurations.  The Euclidean action $I = \wt I + \Delta I$ is the spacetime integral of the Wick-rotated Lagrangian plus boundary terms appropriate to our choice of the canonical ensemble \cite{Gibbons:1976ue}. 
Here, the boundary conditions at asymptotic infinity are defined by periodicity $\beta$ in Euclidean time, with a total electric flux $Q$ at the boundary \cite{Braden:1990hw}, though we suppress all dependence on the latter throughout.

By assumption, the higher-dimension operators in the low-energy effective theory are dominated by tree-level contributions from heavy fields.    The Euclidean path integral including these ultraviolet modes is 
\eq{
 \int d[\hat g]d[ \hat A]d[\hat \phi] \, e^{-I_{\rm UV}[\hat g, \hat A, \hat \phi ]} =\int d[\hat g]d[ \hat A] \, e^{-I[\hat g , \hat A]},
}{eq:UVcomplete}
where $\hat \phi$ is a collective integration variable running over all configurations of the heavy fields.  As a convention, we choose $\hat \phi=0$ as the boundary condition at asymptotic infinity, thus defining zero as the vacuum expectation value of the field in flat space.  In the classical limit, the right-hand side of \Eq{eq:UVcomplete} is obtained by solving the equations of motion for the heavy fields and plugging these solutions back into the action.

Now consider an {\it alternative} field configuration that instead sets all the heavy fields to zero, thus rendering them non-dynamical.  This field configuration does {\it not} satisfy the equations of motion, but this will have no bearing on the following argument.
For this configuration the massive fields are decoupled and their contributions to higher-dimension operators are set strictly to zero.  It is then a simple mathematical fact that
\eq{
I_{\rm UV}[\hat g, \hat A, 0] = \wt I [\hat g,\hat A],
}{eq:plugIUV}
where the right-hand side is the Euclidean action for pure Einstein-Maxwell theory for any choice of metric and gauge field.  The statement of  \Eq{eq:plugIUV} encodes our assumption that the dominant contributions to the higher-dimension operators in the effective field theory come from heavy fields.

Putting this all together, we obtain a simple inequality relating the free energy of the perturbed black hole and a Reissner-Nordstr\"om black hole at the same temperature,\footnote{Strictly speaking, the free energy of a black hole is obtained only after subtracting the free energy contribution from hot flat space or some other reference spacetime \cite{Brown:1992br}.  However, since $\log Z(\beta)$ and $\log \wt Z(\beta)$ have the same asymptotic boundary conditions, any such reference dependence will cancel from either side of \Eq{eq:logZineq}.}
\eq{
-\log Z(\beta) = I_{\rm UV}[g_\beta,A_\beta,\phi_\beta] < I_{\rm UV}[\wt g_\beta, \wt A_\beta, 0] =    \wt I [\wt g_\beta,\wt A_\beta] = -\log \wt Z(\beta).
}{eq:logZineq}
To obtain the first equality we compute $\log Z(\beta)$ via the saddle-point approximation.  Here $g_\beta$, $A_\beta$, and $\phi_\beta$ are the solutions to the classical equations of motion with subscripts to emphasize their consistency with boundary conditions enforcing inverse temperature $\beta$.  By definition, $ I_{\rm UV}[g_\beta,A_\beta, \phi_\beta]$ extremizes the Euclidean action.  The subsequent inequality then holds if this extremum is {\it also} a local minimum, in which case off-shell field configurations slightly displaced from the  classical solutions will increase the Euclidean action.  For this off-shell field configuration we choose the pure Reissner-Nordstr\"om metric $\wt g_\beta$ subject to the same boundary condition dictating inverse temperature $\beta$, while pinning all heavy fields to zero.   Since this configuration differs only marginally from the true solution of the equations of motion, the displacement  from the local minimum will be tiny as required. From \Eq{eq:plugIUV} we see that the resulting expression is formally equal to the Euclidean path integral for pure Einstein-Maxwell theory evaluated on the Reissner-Nordstr\"om background.  Using the saddle-point approximation once more, we obtain the final equality with $-\log \wt Z(\beta)$, which is $\beta$ times the free energy $\widetilde F(\beta)$ for a Reissner-Nordstr\"om black hole at inverse temperature $\beta$.

Crucially, $\log \wt Z(\beta)$ does {\it not} correspond to the free energy of a Reissner-Nordstr\"om black hole of mass $M$, which has an inverse temperature $\wt \beta$.   To relate \Eq{eq:logZineq} to the latter, we plug \Eq{eq:beta_def} into the right-hand side of \Eq{eq:logZineq}, yielding
\eq{
\log \wt Z(\beta) = \log \wt Z(\wt \beta) - M \partial_M  \Delta S,
}{eq:shiftZtoZ}
where $\log \wt Z(\wt \beta)$ is the free energy of a Reissner-Nordstr\"om black hole of mass $M$ and inverse temperature $\wt \beta$.  To obtain \Eq{eq:shiftZtoZ}, we inserted the thermodynamic relation, $M = - \partial_{\tilde \beta} \log \wt Z(\wt \beta)$, together with the formula for the inverse temperature shift, $\Delta \beta = \partial_M \Delta S$.  From the definition of the free energy of the canonical ensemble, we then obtain
\eq{
  \log Z (\beta)&=S - \beta M = (1- M\partial_M)S \\
    \log \wt Z (\wt \beta)&=\wt S - \wt \beta M = (1- M\partial_M)\wt S ,
}{eq:ZviaS}
where we have used the fact that the perturbed black hole at inverse temperature $\beta$ has the same mass as the unperturbed black hole at inverse temperature $\wt \beta$.
Combining Eqs.~\eqref{eq:logZineq}, \eqref{eq:shiftZtoZ}, and \eqref{eq:ZviaS}, we cancel terms to obtain
\eq{
\Delta S >0,
}{eq:dSproof}
establishing our claim.  The above argument accords with the natural intuition that constraining microscopic states, i.e., heavy fields, to be non-dynamical will decrease the entropy.  

Let us comment briefly on a subtle but important caveat to the above arguments.  The inequality in \Eq{eq:logZineq} crucially assumes that on the black hole solution the Euclidean action is not just an extremum but specifically a local minimum.  The latter condition guarantees the stability of the Euclidean action under small off-shell perturbations.
As is well known, however, the Euclidean path integral suffers from saddle-point instabilities mediated by conformal perturbations of the metric that are unbounded from below.  Fortunately, it was shown in Refs.~\cite{Gibbons:1978ac,Gibbons:1978ji} that these particular modes are actually a gauge artifact.  For a certain orthogonal decomposition of the metric, the offending conformal mode can be completely decoupled from the physical degrees of freedom.  With an appropriate contour of integration it is then possible to path integrate over this mode to yield a convergent final expression.

Later on, an analysis of the Euclidean Schwarzschild solution \cite{Gross:1982cv} revealed a bona fide instability coming from a certain non-conformal perturbation about the background solution.  This result has been interpreted as evidence that this solution actually describes a tunneling event from a hot background spacetime into a large black hole \cite{York:1986it,Hawking:1982dh}.  Later analyses \cite{Prestidge:1999uq,Reall:2001ag,Monteiro:2008wr} support these claims and moreover show a direct correlation between the existence of negative modes and the thermodynamic instability that arises from negative specific heat.  To our knowledge, in all cases considered these saddle-point instabilities disappear when the specific heat is positive, which for example in $D=4$ requires a black hole with $q/m > \sqrt{3}/{2}$ in natural units.

For the remainder of this paper we restrict to black holes within this thermodynamically stable window of charge-to-mass ratios so that the extremum of the Euclidean action is a local minimum rather than a saddle point and our proof of \Eq{eq:dSproof} applies. 
Crucially, this range of parameters includes highly charged black holes, so the results of this section can be used in our discussion of the WGC later on.\footnote{The stability results  of Refs.~\cite{Gibbons:1978ac,Gibbons:1978ji,Gross:1982cv,York:1986it,Hawking:1982dh,Prestidge:1999uq,Reall:2001ag,Monteiro:2008wr} were obtained in the context of pure gravity and Einstein-Maxwell theory.  However, they should also apply in the presence of additional heavy fields, which at low energies only produce small corrections to the leading-order black hole solutions.  The precise crossover from positive to negative specific heat may be slightly shifted by the effects of the corresponding higher-dimension operators but this has no impact on the thermodynamic stability of highly charged black holes, which are safely within this window.}

We believe that \Eq{eq:dSproof} is likely true even after relaxing some of the assumptions outlined in \Sec{sec:assumptions}, specifically those requiring a tree-level quantum field theoretic ultraviolet completion.  In particular, from Eqs.~\eqref{eq:shiftZtoZ} and \eqref{eq:ZviaS} it is obvious that the positivity condition in \Eq{eq:dSproof} is mathematically equivalent to an inequality of free energies,
\eq{
F(\beta) < \wt F(\beta),
}{eq:diffF}
which says that the free energy of the perturbed black hole is less than that of a Reissner-Nordstr\"om black hole at the same temperature.  Here we emphasize that the former is computed in the theory defined by $\cal L$ and the latter corresponds to $\wt{\cal L}$.  It is quite possible that the free energy condition in
   \Eq{eq:diffF}  holds in complete generality, e.g., including quantum corrections.  
   
\subsection{Explicit Example}\label{sec:example}

 It is instructive to examine the above arguments for the explicit example of a massive, gravitationally-coupled scalar field.  The Euclidean action for the ultraviolet completion is\footnote{Here we ignore boundary terms since we will be interested only in the low-energy corrections generated from integrating out heavy fields.  These states produce higher-derivative effective operators whose effects fall off quickly with distance and are thus subdominant to bulk  action contributions.}
\be 
\begin{aligned}
I_{\rm UV}[g, A, \phi] = \int d^D x \sqrt{ g}  \bigg[ & -\frac{1}{2\kappa^2} R +\frac{1}{4} F_{\mu\nu} F^{\mu\nu}\\ & + \left(\frac{a_\phi}{\kappa} {R} + b_\phi  \kappa F_{\mu\nu} F^{\mu\nu}\right)\phi + \frac{1}{2} \nabla_\mu\phi \nabla^\mu \phi + \frac{1}{2} m_\phi^2 \phi^2 \bigg],\label{eq:I_UV}
\end{aligned}
\ee
out of notational convenience dropping the hatted convention employed previously.   The coupling constants $a_\phi$ and $b_\phi$ are dimensionless and have indefinite sign.  The classical solution for $\phi$ is
\eq{
\phi = \frac{1}{\nabla^2 - m_\phi^2}\left(\frac{a_\phi}{\kappa} {R} + b_\phi  \kappa F_{\mu\nu} F^{\mu\nu}\right).
}{}
Plugging back into the Euclidean action, we obtain
\eq{
I[g, A] = \int d^D x \sqrt{ g} \left[- \frac{1}{2\kappa^2} R +\frac{1}{4} F_{\mu\nu} F^{\mu\nu} -\frac{1}{2m_\phi^2} \left(\frac{a_\phi}{\kappa} {R} + b_\phi  \kappa F_{\mu\nu} F^{\mu\nu}\right)^2  \right],
}{eq:I_IR}
where all gradient terms from the $\phi$ solution are negligible in the low-energy limit.
Next, consider the Euclidean action for the full theory, given a field configuration where $\phi$ is set strictly to zero.  The resulting expression is  the Euclidean action for pure Einstein-Maxwell theory,
\eq{
I_{\rm UV}[g, A, 0] = \int d^D x \sqrt{ g} \left[ - \frac{1}{2\kappa^2} R +\frac{1}{4} F_{\mu\nu} F^{\mu\nu}  \right] = \wt I[g,A].
}{eq:I_EM}
Putting this all together, we learn that
\eq{
I[g, A] <I[\wt g, \wt A] < \wt I[\wt g, \wt A].
}{}
The first inequality follows because the action is minimized on the solutions to the classical equations of motion for thermodynamically stable black holes.  The second inequality follows because  \Eq{eq:I_IR} differs from \Eq{eq:I_EM} by a negative-definite contribution.  This relation between Euclidean actions then implies \Eq{eq:diffF} in the saddle-point approximation.

\subsection{Unitarity and Monotonicity}

From \Sec{sec:example} it is clear that the entropy inequality $\Delta S > 0$ is very closely related to unitarity.  In particular, the relative signs derived in the previous example hinged critically on the absence of tachyons or ghosts in the ultraviolet completion.   This is not so surprising, since the presence of such pathologies introduce saddle-point instabilities on a general background, be it flat space or a black hole.   It would be interesting to draw a direct connection between our results and previous discussions of unitarity and analyticity \cite{Adams:2006sv,Rattazzi}.

There is also an interesting connection between our results and monotonicity theorems along renormalization group flows \cite{Komargodski:2011vj}.   Our proof of $\Delta S >0$ was framed in terms of integrating out all heavy fields at once.  However, if the spectrum of particles is hierarchical, then this logic can be applied at each mass threshold in sequence.  The total entropy shift is then
\eq{
\Delta S = \int_{\rm UV}^{\rm IR} dS,
}{}
where the differential entropy $dS>0$ contributed by each state is positive.  Extrapolating from this classical result, it is reasonable to conjecture that such a positivity condition persists at the quantum level.  Indeed, as we will see in \Sec{sec:checks}, the renormalization of pure Einstein-Maxwell theory accords with this expectation.

It is known that the quantum entropy corrections computed from the Euclidean path integral are in close relation with the entanglement entropies of the corresponding modes, where the horizon is the entangling surface (see Refs.~\cite{Solodukhin:2011gn,Harlow:2014yka} and references therein).   Since entanglement entropy is intrinsically positive, so too are the quantum entropy corrections, to the extent to which they are equivalent.    We will comment on this connection in more detail in \Sec{sec:ROI}.

\section{Classical vs.~Quantum}  \label{sec:UV}

Up until now we have focused on classical corrections to the entropy, ignoring all loop effects.   As we will see, there exists a regime of black hole masses in which the classical contributions dominate over the quantum.  In this case, $\Delta S >0$ according to the proof presented in the previous section.  In what follows, we estimate and compare the characteristic size of the leading tree-level and loop-level corrections to the black hole entropy.

\subsection{Leading Contributions}

For concreteness, consider a scalar $\phi$ of mass $m_\phi$.  As per the assumptions of the previous section,  we assume that this field has the usual minimal coupling to gravitons but also direct couplings to the curvature and electromagnetic field strength.  Conservatively, we assume that these couplings are at least of gravitational strength, so the interactions go as $\sim \phi R / \kappa$ and $\sim \kappa \phi F^2$.  Here it will be convenient to define a set of rescaled higher-dimension operator coefficients,
\eq{
\e_{1,2,3} = \kappa^2 \d_{1,2,3} ,\qquad \e_{4,5,6} =   \d_{4,5,6} ,\qquad \e_{7,8} = \kappa^{-2} \d_{7,8} ,
}{eq:ds}
which are the dimensionally natural basis in which to express quantities. All of these rescaled coefficients have mass dimension $[\e_i]=-2$.

\medskip

\noindent {\bf Tree Level.} The dominant contributions coming from tree-level $\phi$ particle exchange enter as corrections  to the $R^2$, $RF^2$, and $F^4$ operators of size
\eq{
\delta ( \e_i ) \sim \frac{1}{ m_\phi^2} \quad \textrm{(tree)} .
}{eq:dom_tree}
Here each contribution scales with a factor of $1/m_\phi^2$ coming from the $\phi$ propagator denominator. 

\medskip

\noindent {\bf Loop Level.} First, let us consider loop corrections involving purely gravitational interactions.  At one loop, the leading contributions enter through the renormalization of the gravitational constant,
\eq{
\delta (\kappa^{-2}) \sim m_\phi^{D-2} \quad \textrm{(loop)},
}{eq:dom_loop}
which follows straightforwardly from dimensional analysis.  At loop level, gravitational interactions also yield contributions to higher-dimension operators,
\eq{
\delta ( \e_i ) \sim \kappa^2 m_\phi^{D-4} \quad \textrm{(loop)},
}{eq:subdom_loop}
which are always subdominant to \Eq{eq:dom_tree}. 

Loops involving gauge interactions will similarly renormalize the gauge coupling as well as the higher-dimension operator coefficients, yielding contributions that scale as  \Eq{eq:dom_loop} and \Eq{eq:subdom_loop} but with enhancement factors proportional to the charge-to-mass ratios of fundamental charged particles.  In principle, these contributions can dominate.  For example, in the standard model, the leading contributions to the Euler-Heisenberg Lagrangian come from loops of electrons.  However, as shown in \Ref{Cheung:2014ega}, this only happens when there are fundamental charged particles with large charge-to-mass ratios.  In this case there is no claim to prove, since WGC is already satisfied.  For this reason, we restrict our consideration to theories where all fundamental charged particles fail or are near the WGC bound without satisfying it.  In this limit gauge interactions are of the same strength as gravitational interactions so the leading tree-level and loop-level corrections from both scale as in  \Eq{eq:dom_tree} and \Eq{eq:dom_loop}.

\subsection{Region of Interest} \label{sec:ROI}

From Eqs.~\eqref{eq:dom_tree}, \eqref{eq:dom_loop}, and \eqref{eq:subdom_loop}, we can estimate the corresponding corrections to the black hole entropy, which takes the schematic form
\eq{
S \sim \frac{\rho^{D-2}}{\kappa^2} +\rho^{D-2} m_\phi^{D-2} + \rho^{D-4} m_\phi^{D-4}+ \frac{\rho^{D-4}}{\kappa^2 m_\phi^2} +\cdots,
}{eq:S_schematic}
where $\rho$ is the radius of the black hole.
The first term is the Bekenstein-Hawking entropy, the second term is the quantum correction to the gravitational constant, and the third and fourth terms are the quantum and classical corrections to the higher dimension operators, respectively.  Demanding that classical entropy corrections dominate over all quantum corrections requires that
\eq{
\rho \ll \frac{1}{\kappa m_\phi^{D/2}}.
}{eq:rHbound}
Crucially, since $m_\phi$ is much smaller than the Planck scale, this constraint is consistent with $\rho \gg 1/m_\phi$, which is necessary to remain within the regime of validity of the effective field theory.  For the remainder of this paper we focus on this regime of black hole masses.

Before moving on, let us comment briefly on the expectation of positivity for the quantum entropy corrections. While our results only rely on positivity of the classical contribution, it is reasonable to conjecture that the same might apply to quantum corrections.  It is known, however, that the quantum contributions in \Eq{eq:subdom_loop} have indefinite sign and in $D=4$ these correspond to well-studied logarithmic corrections to black hole entropy \cite{Solodukhin:1994yz,Solodukhin:1994st,Fursaev:1994te,Carlip:2000nv}.  Nevertheless, these signs do not matter because we have already shown that these corrections are parametrically subdominant to the contributions from \Eq{eq:dom_loop} related to the renormalization of the gravitational constant.   

Meanwhile, corrections of the latter type have also been computed via heat kernel methods and found to be positive for minimally-coupled spin 0 and 1/2 particles but negative for spin 0 particles with non-minimal couplings \cite{Solodukhin:1995ak} as well as spin 1 and spin 2 particles \cite{Kabat:1995eq,Solodukhin:2015hma}.  There is, however, a longstanding debate over the physical meaning of these negative corrections.  They indicate a naive mismatch with calculations of quantum field theoretic entanglement entropy, which is manifestly positive.  While these contributions have been understood as the entanglement of certain edge modes \cite{Donnelly:2012st,Donnelly:2014fua}, the sign of the leading power-law divergence was also shown to be regulator-dependent.   In general, these negative corrections are formally power-law divergent and scheme-dependent and such quantities should be at least partly absorbed into the renormalized gravitational constant \cite{Susskind:1994sm,Solodukhin:1994yz,Demers:1995dq,Solodukhin:2015hma}.  For a consistent ultraviolet completion, all divergences will disappear and the residual corrections will be finite.  On physical grounds, it is expected that, if properly regulated, these quantum entropy corrections will be manifestly positive as expected from the manifest positivity of entanglement entropy.

\section{Black Hole Spacetime} \label{sec:spacetime}

We now turn to the study of a spherically symmetric, positively charged black hole of mass $M$ and charge $Q$ in the presence of low-energy corrections to pure Einstein-Maxwell theory.  For simplicity, we restrict to $D=4$ dimensions for the remainder of the body of this paper, but all of our results generalize to arbitrary spacetime dimension $D \geq 4$, as shown in \App{app:genD}.

Our aim is to derive new bounds on the higher-dimension operator coefficients $\d_i$.  As noted previously, this restricts our consideration to black holes large enough that the effective field theory is valid but small enough to satisfy \Eq{eq:rHbound}, so $1/m_\phi \ll \rho \ll 1/\kappa m_\phi^{2}$, where  $ m_\phi$ is the mass scale of the new states.  This range always exists provided there is a parametric separation between $ m_\phi$ and the Planck scale.     Furthermore, we consider the thermodynamically stable regime where $q/m > \sqrt{3}/{2}$ so the specific heat is positive.

Note that the mass and charge are defined at spatial infinity.   Since we will only consider static spacetimes, the ADM and Komar formulations of these quantities are equivalent. Explicitly, the Komar mass and charge are\footnote{Here we take all black holes to be of definite mass even though in practice they have a small width given by their inverse lifetime.  For instance, as classically stable objects, black holes can only decay quantum mechanically via Hawking emission and Schwinger pair production processes, which are suppressed by additional factors of the gravitational constant.  Moreover, decays into smaller black holes will proceed via non-perturbative gravitational effects, which are exponentially suppressed.}
\eq{
M \sim \frac{1}{\kappa^2} \int_{i^0} n_\mu \sigma_\nu \nabla^\mu K^\nu   \qquad {\rm and} \qquad   Q \sim \int_{i^0}  n_\mu \nabla_\nu F^{\mu\nu}  ,}{}
where the integral region $i^0$ is spatial infinity, $n$ is the unit timelike normal vector, $\sigma$ is the unit spatial outward-pointing normal vector, and $K$ is the timelike Killing vector. Since the integral is evaluated at spatial infinity, only the leading behavior at large $r$ contributes to these expressions.   Because higher-dimension operators correct the metric and gauge field at subleading order in $r$, they do not affect the definition of the asymptotic mass and charge.

\subsection{Unperturbed Solution}

The unperturbed theory is described by the Lagrangian for Einstein-Maxwell theory in \Eq{eq:EM}.  For notational convenience we will sometimes describe the mass and charge in natural units of the gravitational constant,\footnote{In $D$ dimensions, the mass dimensions of various quantities are $[\kappa^2]=2-D$, $[R]=2$, $[F]=D/2$, $[M]=1$, $[Q]=2-D/2$, $[m]=[q]=3-D$, 
$[\d_{1,2,3}]=D-4$, $[\d_{4,5,6}]=-2$, $[\d_{7,8}]=-D$, and $[\e_i] = -2$. \label{foot:massdim}}
\eq{
m=\frac{\kappa^2 M}{8\pi} \qquad {\rm and} \qquad q= \frac{\kappa Q}{4\sqrt{2}\pi}.
}{eq:units}
We also define the extremality parameter
\eq{
\qquad \xi = \sqrt{1-\frac{q^2}{m^2}},
}{eq:xidef}
where a neutral black hole corresponds to $\xi=1$ and an extremal black hole corresponds to $\xi=0$.   As noted previously, we will consider black holes with positive specific heat, corresponding to $q/m > \sqrt{3}/{2}$, or equivalently, $\xi <1/2$.

The solution is the Reissner-Nordstr\"om black hole, whose metric takes the static and spherically symmetric form
\eq{
{\rm d}s^2 = \wt g_{\mu\nu} {\rm d}x^\mu {\rm d}x^\nu =  -\wt f(r)\mathrm{d}t^2 + \frac{1}{\wt g(r)}{\rm d}r^2 + r^2 {\rm d}\Omega^2,
}{eq:sphericalmetric}
where the unperturbed metric components are
\eq{
\wt f(r)=\wt g(r) =1-\frac{2m}{r}+\frac{q^2}{r^2}
}{eq:fg0}
and the unperturbed electromagnetic field strength is
\eq{
\wt F_{\mu\nu} {\rm d}x^\mu \wedge {\rm d}x^\nu=\frac{Q}{4\pi {{r}^{2}}} {\rm d}t \wedge {\rm d}r.
}{eq:Fbkg}
The unperturbed event horizon is the outer horizon of the Reissner-Nordstr\"om black hole and is located at the radius  $r=\wt \rho$, where
\eq{
\wt \rho  = m + \sqrt{m^2 -q^2} = m(1+\xi).
}{eq:rH0}
The absence of a naked singularity implies that the charge is bounded by the inequality
\eq{
\frac{q}{m} \leq 1,
}{}
which is saturated in the case of an extremal black hole.

\subsection{Perturbed Solution}

In the presence of the higher-dimension operators in \Eq{eq:L}, the perturbed metric takes the form
\eq{
{\rm d}s^2 =   g_{\mu\nu} {\rm d}x^\mu {\rm d}x^\nu =  -   f(r)\mathrm{d}t^2 + \frac{1}{  g(r)}{\rm d}r^2 + r^2 {\rm d}\Omega^2,
}{eq:sphericalmetricperturbed}
where the metric components are complicated functions of the coefficients $\d_i$.
However, as shown in Refs.~\cite{Kats:2006xp,Campanelli:1994sj}, it is straightforward to compute corrections to the Reissner-Nordstr\"om solution order-by-order in $\d_i$.  Following the prescription in \Ref{Kats:2006xp} applied to the higher-dimension operators in \Eq{eq:L}, we find that at first order in $\d_i$ the radial component of the metric is
\eq{
g(r)&=1-\frac{2m}{r}+\frac{q^2}{r^2} -\frac{q^2}{r^6}\left\{
\begin{array}{l}
\phantom{+} \dfrac{4}{5} (\e_2+4\e_3) (6q^2 - 15mr+10r^2)\\
+8\e_4 (3q^2 -7mr+4r^2)+\dfrac{4}{5}\e_5 (11q^2 - 25 mr+15r^2)\\
 +\dfrac{4}{5}\e_6(16q^2 - 35mr+20r^2)+\dfrac{8}{5}(2\e_7 + \e_8)q^2
 \end{array} \right\},
}{eq:g}
where the coefficients $\e_i$ are defined in terms of $\d_i$ in \Eq{eq:ds}.

\section{Calculation of Entropy}\label{sec:entropy}

\subsection{Wald Entropy Formula}

We now compute the entropy corrections to black holes of size much greater than the Compton wavelengths of the heavy modes.  A major advantage of this approach is that the effects of all short-distance degrees of freedom are  encoded purely by higher-dimension operators.  Moreover, even though these states are absent from the low-energy theory, their contributions to the entropy are fully accounted for by the Wald formula in \Eq{eq:Waldgeneral}.  

For a spherically symmetric spacetime, the integral in \Eq{eq:Waldgeneral} is trivial, yielding
\eq{
S=-2\pi  A \, \frac{\delta {\mathcal{L}}}{\delta {{R}_{\mu\nu\rho\sigma}}} \, {{ \epsilon }_{\mu\nu}}{{ \epsilon }_{\rho\sigma}} \bigg|_{ g_{\mu\nu}, \, \rho},
}{}
where all quantities are evaluated for the perturbed metric and perturbed horizon radius, $\rho = \wt\rho +\Delta \rho$.  The perturbed horizon area is $A = 4\pi \rho^2 $ and the binormal is
\eq{
 \epsilon_{\mu\nu}(r) = \sqrt{\frac{ f(r)}{ g(r)}}(\delta_\mu^t \delta_\nu^r - \delta_\mu^r \delta_\nu^t).
}{}
Expanding the area $A = \wt A + \Delta A$ and the Lagrangian $\LL = \wt \LL + \Delta \LL$ in perturbations, we obtain
\eq{
S=-2\pi  \left(
\wt A \, \frac{\delta \wt  {\mathcal{L}}}{\delta {{R}_{\mu\nu\rho\sigma}}}
+
\wt A \, \frac{\delta \Delta {\mathcal{L}}}{\delta {{R}_{\mu\nu\rho\sigma}}}
+
\Delta A \, \frac{\delta  \wt{\mathcal{L}}}{\delta {{R}_{\mu\nu\rho\sigma}}}   + \cdots
 \right){{\epsilon }_{\mu\nu}}{{\epsilon }_{\rho\sigma}} \bigg|_{\, g_{\mu\nu} , \, \rho} ,
}{eq:Sexpand}
where $\wt A = 4\pi  \wt\rho^2 $ and the ellipses denote terms that are higher order in perturbations. The first term is straightforwardly obtained
by differentiating \Eq{eq:EM} with respect to the Riemann tensor,
\eq{
\frac{\delta  \wt{\mathcal{L}}}{\delta R_{\mu\nu\rho\sigma}} = \frac{1}{2\kappa^2}g^{\mu\rho}g^{\nu\sigma},
}{eq:dLdR0}
where the proper (anti)symmetrization of indices on the right-hand side is implicit.
Since the binormal is normalized as $\epsilon_{\mu\nu}\epsilon^{\mu\nu}=-2$, the first term in \Eq{eq:Sexpand} is simply the unperturbed black hole entropy $\wt S$ defined in \Eq{eq:BHentropy}.  Moving this term to the left-hand side of \Eq{eq:Sexpand}, we obtain an expression for the difference in entropies,
\eq{
\Delta S =  \Delta S_{\rm I} + \Delta S_{  \rm H} ,
}{eq:S}
split into an ``interaction'' and ``horizon'' contribution.  Because we are working at first order in perturbations, both of these terms should be evaluated on the unperturbed metric and horizon radius. The interaction contribution,
\eq{
\Delta S_{\rm I} =-2\pi    \wt A \, \frac{\delta \Delta {\mathcal{L}}}{\delta {{R}_{\mu\nu\rho\sigma}}}   {{\epsilon }_{\mu\nu}}{{\epsilon }_{\rho\sigma}}  \bigg|_{\, \wt g_{\mu\nu} , \,  \wt\rho} ,
}{eq:Sdyn}
appears because the interactions of photons and gravitons are modified at low energies.  Meanwhile, the horizon contribution,
\eq{
\Delta S_{\rm H} =  -2\pi   \Delta A \, \frac{\delta \wt {\mathcal{L}}}{\delta {{R}_{\mu\nu\rho\sigma}}}  {{\epsilon }_{\mu\nu}}{{\epsilon }_{\rho\sigma}}  \bigg|_{\, \wt g_{\mu\nu} , \,  \wt\rho} = \frac{2\pi}{\kappa^2} \Delta A,
}{eq:Sgeom}
is present because higher-dimension operators modify the black hole background, thus shifting the location of the horizon.  Here we have substituted in \Eq{eq:dLdR0} to write the right-hand side of this expression as simply the shift of the horizon area.

\subsection{Interaction Contribution}

To obtain the interaction contribution to the entropy shift we compute
\eq{
\frac{\delta \Delta \mathcal{L}}{\delta R_{\mu\nu\rho\sigma}} =& \phantom{{}+{}} 2\d_1 R g^{\mu\rho}g^{\nu\sigma} + 2\d_2 R^{\mu\rho}g^{\nu\sigma}+2\d_3 R^{\mu\nu\rho\sigma}\\&+\d_4 F_{\alpha\beta}F^{\alpha\beta}g^{\mu\rho}g^{\nu\sigma} +\d_5 F^\mu_{\;\;\;\alpha}F^{\rho\alpha}g^{\nu\sigma} + \d_6 F^{\mu\nu}F^{\rho\sigma},
}{eq:Waldpre}
where proper (anti)symmetrization on indices is left implicit as before.
Substituting the unperturbed black hole background into \Eq{eq:Waldpre} and evaluating \Eq{eq:Sdyn}, we obtain
\eq{
\Delta S_{\rm I} = \wt S\times \frac{2}{m^2(1+\xi)^3}\left[8\e_3 -2(1-\xi)(\e_2 + 6\e_3 + 2\e_4 + \e_5 + 2\e_6)\right],}{eq:SdynCALC}
written in terms of the coefficients defined in \Eq{eq:ds}. Setting  $\xi = 1$ in our expression for $\Delta S_{\rm I}$ in \Eq{eq:SdynCALC} agrees with the expressions in \Ref{Goon} as well as their generalization to arbitrary dimension in \Ref{Clunan:2004tb}.

\subsection{Horizon Contribution}\label{sec:HC}

The horizon contribution to the entropy shift depends on the location of the perturbed horizon.    Since the spacetime is static, the horizon is determined by zeros of the metric components $f(r)$ and $g(r)$ defined in \Eq{eq:sphericalmetricperturbed}.  On general grounds, $f(r)$ and $g(r)$ have coincident zeros since otherwise the spacetime would contain a region with non-Lorentzian signature.  Moreover, we have verified by explicit calculation that $f(r)$ and $g(r)$ share the same zeros at first order in perturbations.

The perturbed horizon is located at radius $\rho = \wt\rho + \Delta\rho $.  To compute $\rho$, we expand $g(r)= \wt g(r) + \Delta g(r)$ at first order in perturbations, as defined in \Eq{eq:g}.  The perturbed horizon radius then satisfies the equation
\eq{
0 = g(\rho) = \wt g(\wt\rho) + \Delta g(\wt\rho) +  \Delta\rho \,  \partial_{\tilde \rho}  \wt g(\wt\rho).
}{eq:horizoncondition}
The first term on the right-hand side vanishes by the definition of the unperturbed horizon radius.  Solving for the horizon shift, we find
\eq{
\Delta\rho = -  \frac{\Delta g(\wt\rho)}{   \partial_{\tilde \rho}  \wt g(\wt\rho)  }.
}{eq:delta_rho}
At first order, the perturbed horizon area is then given by
\eq{
\Delta A =A- \wt A= 8\pi  \wt\rho \Delta\rho =  -\frac{8\pi \wt\rho \Delta g(\wt\rho)}{ \partial_{\tilde \rho}  \wt g(\wt\rho)}.
}{eq:dAgeneral}
Inputting the perturbed metric in \Eq{eq:g} and evaluating \Eq{eq:Sgeom}, we obtain
\eq{
\Delta S_{\rm H}  ={\wt S}\times  \frac{4(1-\xi)}{5  m^2 \xi (1+\xi)^3}[
(1+4\xi)(\e_2 + 4\e_3+ \e_5 + \e_6) + 10\xi \e_4 + 2(1-\xi)(2\e_7 + \e_8)].
}{eq:SgeomCALC}
Note that the horizon contribution to the entropy shift is divergent in the  $\xi \rightarrow  0$ limit corresponding to an extremal black hole.   Physically, this occurs because the inner and outer horizons become degenerate, so $ {\partial_ {\tilde \rho}  \wt g(\wt\rho)}\rightarrow 0$.   In this case, \Eq{eq:horizoncondition} implies that for some fixed contribution $\Delta g(\wt\rho)$ from higher-dimension operators, the horizon must shift by a parametrically large amount $\Delta\rho$ in order to maintain the horizon condition.

Of course, the strict $\xi\rightarrow 0$ limit is pathological since this produces an infinite entropy shift, signaling a breakdown of perturbation theory.   Demanding that the shift in entropy be much smaller than the unperturbed entropy, we obtain the constraint
\eq{
\xi \gg \frac{|\e_i|}{m^2}.
}{eq:perturbativitycondition}
As shown in \Eq{eq:rHbound}, the classical effects of higher-dimension operators are only dominant over the quantum for black hole radii smaller than a certain value, $\rho \ll 1/\kappa m_\phi^2$.  Using that $\e_i \sim 1/ m_\phi^2$ for a tree-level ultraviolet completion, \Eq{eq:perturbativitycondition} becomes $\xi \gg \kappa^2  m_\phi^2$. 

Perturbativity also requires that the shift in the inverse temperature $\Delta \beta$ be subdominant to the background inverse temperature $\wt \beta$ of the unperturbed Reissner-Nordstr\"om black hole.  As a consistency check, we have verified that $\beta = \partial_M S$ agrees with the surface gravity of the perturbed black hole metric.  For highly charged black holes, the unperturbed inverse temperature goes as $\wt \beta \sim m/\xi$ while $\Delta \beta \sim \e_i/m\xi^3$.  Demanding that the correction be smaller than the leading contribution, $\Delta \beta \ll \wt \beta$, we obtain the stronger condition
\be
\xi \gg \frac{{|\e_i|^{1/2}}}{m}. 
\ee
Combining this with the upper bound on $\rho$ from \Eq{eq:rHbound} and the scaling of $\e_i$, we obtain
\be
\xi \gg \kappa m_\phi.
\ee
Thus,  it is always possible to take the limit of a highly charged black hole, $\xi \ll 1$, provided the mass scale $m_\phi$ of the heavy fields is far below the Planck scale as we have assumed throughout.

\section{New Positivity Bounds}\label{sec:bound}

\subsection{General Bounds}
\label{sec:general_bounds}

The total entropy shift $\Delta S = \Delta S_{\rm I} + \Delta S_{\rm H}$ is obtained by adding
 \Eq{eq:SdynCALC} and \Eq{eq:SgeomCALC}, yielding
\eq{
{\Delta S} =& {\wt S}\times \frac{4}{5 m^2 \xi (1+\xi)^3}\times \\&\;\, \times \left[(1-\xi)^2(\e_2 + \e_5) + 2(2+\xi+7\xi^2)\e_3   + (1-\xi)(1-6\xi)\e_6 + 2(1-\xi)^2(2\e_7+\e_8)\right].
}{eq:SIRfinal}
As proven in \Sec{sec:proof}, the entropy shift is positive under the assumptions we have stated.  Combining \Eq{eq:inequality} and \Eq{eq:SIRfinal}, we obtain a family of positivity bounds,
\eq{
(1-\xi)^2 \e_0 +20 \xi \e_3 - 5 \xi(1-\xi) (2\e_3 + \e_6) > 0,
}{eq:bound4D}
where we have defined the parameter
\eq{
\e_0 = \e_2 + 4\e_3 + \e_5 + \e_6 + 4\e_7 + 2 \e_8.
}{eq:d0def}
The bound in \Eq{eq:bound4D} is the main result of this work: a consistency condition on the coefficients of higher-dimension operator corrections to Einstein-Maxwell theory following from the positivity of corrections to the black hole entropy.  The generalizations of Eqs.~\eqref{eq:SIRfinal}, \eqref{eq:bound4D}, and \eqref{eq:d0def} to arbitrary dimension $D$ are derived and presented in Eqs.~\eqref{eq:dSD}, \eqref{eq:boundD}, and \eqref{eq:xyzD} in \App{app:genD}.

\begin{figure}[t]
\begin{center}
\hspace{-5mm} \includegraphics[height=0.45\textwidth]{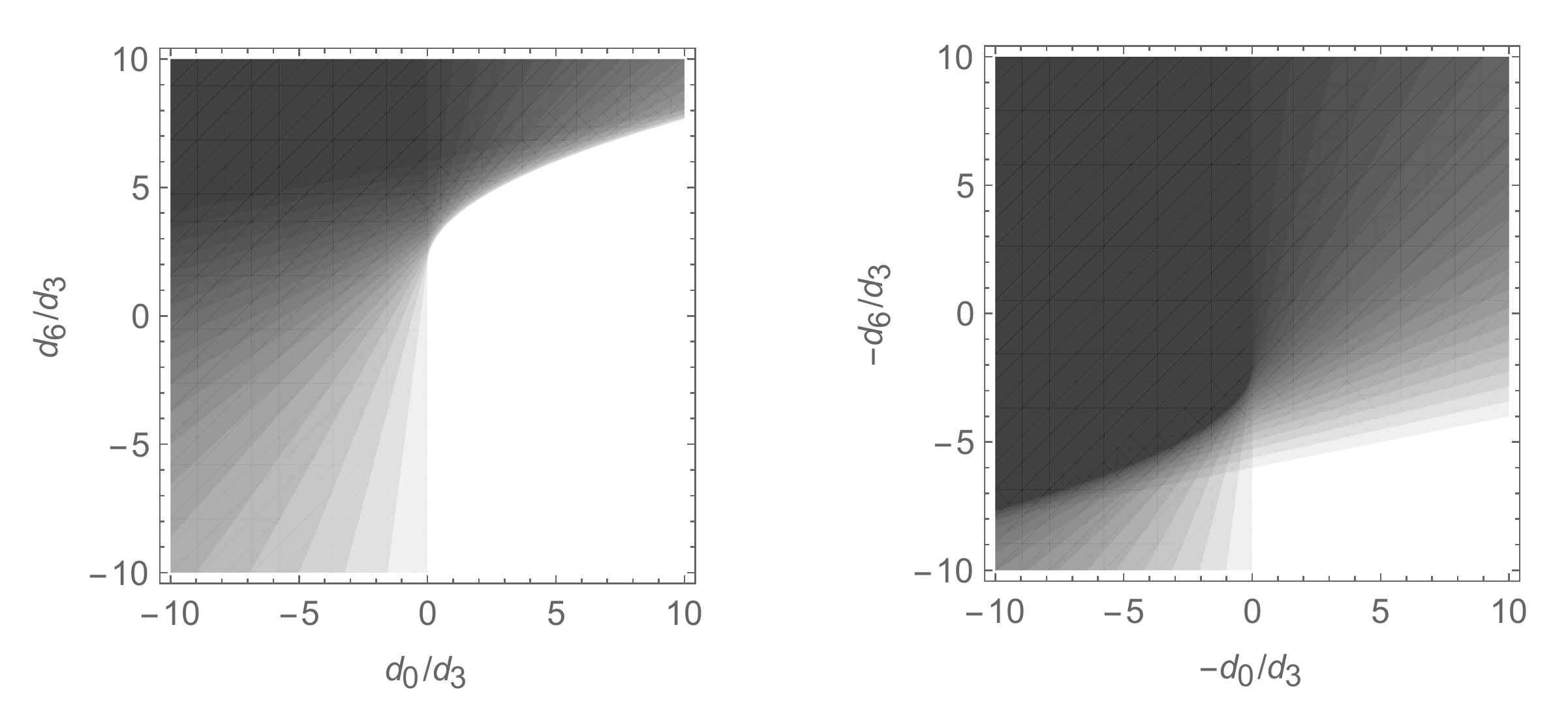}
\end{center}
\vspace{-5mm}
\caption{Constraints on higher-dimension operator coefficients derived from black hole entropy.  The shaded regions are excluded, with the gradations corresponding to incremental values of $\xi \in (0,1/2)$.  The left and right panels correspond to $\e_3>0$ and $\e_3 <0$, respectively.  In either case, $\e_0<0$ is forbidden so $\e_0 >0$ and the WGC is automatically satisfied.  }
\label{fig:bounds}
\end{figure}

As discussed in \Sec{sec:proof}, our proof of $\Delta S >0$ applies to thermodynamically stable black holes, restricting consideration to the window $\xi \in (0,1/2)$.
The full space of bounds over this range defines a convex region in the space of coefficients $\e_i$.  Thus the full space of positivity constraints, depicted in \Fig{fig:bounds} for $D=4$, is stronger than those implied by any finite set of choices for $\xi$.  

We have derived these positivity conditions from a particular physical setup: a black hole of a given charge-to-mass ratio corresponding to $\xi \in (0,1/2)$ and mass consistent with \Eq{eq:rHbound} so that our proof of $\Delta S >0$ applies.  Despite this specificity, we emphasize that the resulting bounds in \Eq{eq:bound4D} are consistency conditions on the effective action and thus hold {\it independently of the background}.

\subsection{Examples and Consistency Checks} \label{sec:checks}

Our bound in \Eq{eq:bound4D} and its higher-dimensional generalization in \Eq{eq:boundD}  pass a number of explicit checks.    To begin, we emphasize that these inequalities are invariant under a change of the field basis, which by reparameterization theorems should leave physical observables unchanged.  As discussed in \App{app:FD}, metric field redefinitions shift the higher-dimension operator coefficients $\e_i$ in a way that renders individual coefficients operationally meaningless.  By studying these transformations, one can build a basis of field redefinition invariant combinations of coefficients, of which $\e_0, \e_3, \e_6$ are a subset.  Remarkably, while the separate contributions to the entropy from $\Delta S_{\rm I}$ and $\Delta S_{\rm H}$ are {\it not} invariant under field redefinitions, their sum $\Delta S$ is  invariant for all $\xi$ in arbitrary dimension $D$, as shown in \App{app:FD}.  Field redefinition invariance of the inequality $\Delta S >0$ is a prerequisite for this bound to have physical meaning.
 
 On physical grounds, it is natural to expect that any positivity condition is preserved under renormalization group flow into the infrared.  This is true because a consistent theory should continue to be consistent at arbitrarily long distances.  Interestingly, this expectation agrees with known results on the one-loop divergences of pure Einstein-Maxwell theory in $D=4$, which enter solely through the $R_{\mu\nu}R^{\mu\nu}$ operator \cite{Deser:1974cz}.  The sign of this divergence is consistent with a negative beta function, indicating that the coefficient of this operator indeed increases in the infrared, consistent with \Eq{eq:bound4D}.

We can also study \Eq{eq:boundD} in simple concrete examples.   First, consider any theory in general dimension $D$ in which the strength of gravity is negligible relative to gauge interactions.   In this limit all higher-dimension operators are vanishing except for $\e_7$ and $\e_8$, which control the leading contributions to photon self-interactions in the Euler-Heisenberg effective action.   Applying the logic of \Ref{Adams:2006sv}, we have computed the four-photon scattering amplitude at low energies and found that dispersion relations imply the positivity conditions $2\e_7 + \e_8 > 0$ and $\e_8 >0$, corresponding to different choices of external polarizations.  In this case, the former inequality exactly implies that $\e_0 >0$ in general dimension $D$, thus providing a consistency check of \Eqs{eq:bound4D}{eq:boundD}.

 Second, we examine the scalar model described in \Sec{sec:example}.  Translating from Euclidean to Lorentzian signature, we obtain the higher-dimension operator coefficients
 \eq{
 \e_i = \frac{1}{2m_\phi^2} \times \left\{ a_\phi^2,0,0,2a_\phi b_\phi,0,0,b_\phi^2,0 \right\},
 }{}
which, written in terms of $\e_0$ using \Eq{eq:xyzD}, give
\eq{
\e_0 = \frac{D-3}{8m_\phi^2}\left[(D-4)a_\phi+2(D-2)b_\phi\right]^2,
}{}
which is a perfect square, so the bound in \Eq{eq:boundD} is again satisfied.

Third, we study the low-energy description of the heterotic string, for which the higher-dimension operators have coefficients as given in Refs.~\cite{Gross:1986mw,Kats:2006xp},
\eq{
\e_i = \frac{\alpha'}{64} \times \left\{ 4,-16,4,0,0,0,-3,12\right\},
}{}
where we have absorbed the dependence on the dilaton expectation value into $\alpha'$. Plugging these parameters into \Eq{eq:boundD} yields $(6D^2 - 30D+37)\xi^2 + 2(D-2)\xi + 2D-5>0$, which holds for all $\xi\in(0,1)$ and $D>3$.  Thus, we find that \Eq{eq:boundD} is actually satisfied even beyond the range of thermodynamic stability and the critical dimension of the string.

\subsection{The Weak Gravity Conjecture}\label{sec:WGC}

In the limit of a highly charged black hole we take $\xi \ll 1$ and our general bound in \Eq{eq:bound4D} becomes
\eq{
\e_0 >0,
}{eq:xi04D}
with $\e_0$ defined in terms of the other $\e_i$ in \Eq{eq:d0def}.  

As it turns out, this inequality is intimately connected with the extremality condition for a black hole.
To see why, consider the unperturbed Reissner-Nordstr\"om solution, for which the extremal charge-to-mass ratio is $\wt z =q/m=1$. In general, quantum corrections to the gravitational constant and electric charge will renormalize the right-hand side of this condition. Since these contributions affect black holes of all masses universally, their effects can simply be absorbed into the definitions of mass and charge.  Meanwhile, higher-dimension operators also shift the maximum charge-to-mass ratio permitted for a physical black hole, i.e., a black hole free from naked singularities.   In contrast,  these corrections are mass-dependent, so they induce a physical shift of the extremality condition to $z = \wt z + \Delta z$ \cite{Kats:2006xp}.  

To compute this shift, we analyze the metric component $g(r,z)$, interpreted as a function of both the radius and the charge-to-mass ratio.  The shifted horizon is defined by the condition $g(\rho,z) =0$.    At linear order in perturbations, this condition becomes
\eq{
0 = g(\rho,z) = \wt g(\wt\rho,\wt z) + \Delta g(\wt\rho,\wt z) +  \Delta\rho \,  \partial_{\tilde \rho}  \wt g(\wt\rho,\wt z) +  \Delta z \, \partial_{\tilde z}  \wt g(\wt\rho,\wt z).
}{eq:gshift}
Since the unperturbed black hole is extremal, the first and third terms on the right-hand side are zero.  Solving for the shift in the charge-to-mass ratio yields
\eq{
\Delta z =  -\frac{\Delta g(\wt \rho ,\wt z)}{  \partial_{\tilde z}  \wt g(\wt\rho,\wt z)}.
}{eq:Deltazpre}
By explicit calculation, the charge-to-mass ratio shift is
\eq{
\Delta z  = \frac{2\e_0}{5m^2} > 0,
}{eq:Deltaz}
which is  positive according to \Eq{eq:xi04D}.

As shown in \Ref{Kats:2006xp}, if $\Delta z$ is positive then small black holes {\it automatically} satisfy the WGC \cite{ArkaniHamed:2006dz}, which posits that an Abelian gauge theory consistently coupled to quantum gravity must contain a state  with charge-to-mass ratio greater than unity.  In its original formulation \cite{ArkaniHamed:2006dz}, the WGC was presented with several compelling justifications.  This included overwhelming circumstantial evidence from a long list of explicit examples in quantum field theory and string theory.  A more direct argument was also presented in the form of an elegant thought experiment involving stable black hole remnants \cite{Susskind:1995da,Giddings:1992hh,tHooft:1993dmi,Bousso:2002ju,Banks:2006mm}. 
In particular, due to mass and charge conservation, a charged black hole is stable unless there exist lighter states with a higher charge-to-mass ratio, into which the black hole can decay.  In the infinite-mass limit, the charge-to-mass ratio of an extremal black hole is dictated by the Reissner-Nordstr\"om solution and approaches unity.  Violation of the WGC would then imply the existence of an infinite tower of stable remnants labeled by the extremal black hole mass, asserted as pathological in  \Ref{ArkaniHamed:2006dz}.  While this thought experiment offers some crucial physical intuition for the WGC, it falls short since there do not exist formal proofs that stable black hole remnants are actually inconsistent with more established principles like the covariant entropy bound.  In many cases the number of black hole remnants is very large but still finite; furthermore, the states are labeled by distinct charges and are thus in principle distinguishable \cite{Shiu}.

On the other hand, if the WGC is satisfied then extremal black holes are unstable to decay.  We have shown here that considerations of black hole entropy imply that the charge-to-mass ratio of an extremal black hole increases with decreasing size.  In particular, higher-dimension operators induce a positive shift of the extremality bound, but these corrections decouple for large black hole masses.    The upshot is then that an extremal black hole of a given mass can always decay into smaller extremal black holes of a greater charge-to-mass ratio, following the upper curve in  \Fig{fig:mass_plot}.  Our bound in \Eq{eq:xi04D}---and thus our proof of the WGC---generalizes to $D$ spacetime dimensions, as shown in \App{app:genD}.

Let us comment on the relation between our results and previous work connecting black holes and the WGC.  First, while the argument in this paper makes critical use of extremal black holes, our reasoning is completely different from the original proposal of \Ref{ArkaniHamed:2006dz}, hinging instead on the thermodynamic entropy of black holes rather than their stability.  
More recently, the WGC has also been linked to the cosmic censorship conjecture  \cite{Horowitz:2016ezu}.  We leave an analysis of this and its relationship to black hole entropy for future work.

In other recent studies \cite{Shiu,Fisher}, the WGC has also been evaluated in the context of black hole entropy using methodologies that differ substantially from our own.  Both \Ref{Shiu} and \Ref{Fisher} examine the leading logarithmic corrections to black hole entropy due to the quantum effects of light matter particles.  Such effects are relevant for black holes of size parametrically smaller than the Compton wavelength of the matter.   We, in contrast, consider the opposite regime, which effectively corresponds to a gapped spectrum.  

Furthermore, Refs.~\cite{Shiu,Fisher} argue for the inconsistency of WGC violation through quite different means:  \Ref{Shiu} makes the argument through the appearance of a low cutoff, while  \Ref{Fisher} employs the second law of thermodynamics.   It is crucial to note that our assertion of a positive entropy shift is logically distinct from the second law of thermodynamics, which applies to the difference in entropy before and after a physical process but within the same physical system.  Our construction is instead based on the positivity of classical entropy corrections proven in \Sec{sec:proof}.

Finally,  Refs.~\cite{Shiu,Fisher} and also another interesting approach \cite{Hod} all consider concrete models with explicit spectra of charged and neutral scalars and fermions.  For this reason, these works at best show that certain WGC-violating theories are inconsistent.  This leaves the logical possibility that more complicated theories that violate the WGC might still be judged valid by their analyses.  In comparison, our work applies to large black holes in a general low-energy effective theory, which is insensitive to the precise details of the spectrum and hence constitutes a model-independent argument for the WGC.

\subsection{Entropy, Area, and Extremality}

A priori, it is somewhat miraculous that the entropy constraint in \Eq{eq:xi04D} is literally equivalent to the extremality condition in \Eq{eq:Deltaz}.  To briefly summarize, we have shown that
\eq{
\Delta S \sim \Delta \rho \sim \Delta z > 0,
}{}
so the low-energy corrections to the near-extremal black hole entropy, area, and extremality condition are all proportional to each other and all positive.

Why does the same combination of coefficients $\e_0$ appear in all of these inequalities?  As it turns out, this connection is not so mysterious once one considers the perturbed metric component $g(\rho,z)$ in \Eq{eq:gshift} as a function of the shift in horizon radius $\Delta \rho$ and the shift in the charge-to-mass ratio $\Delta z$.  For a near-extremal black hole of fixed charge and mass, we set $\Delta z=0$ and thus $ \Delta\rho  = - \Delta g/\partial_{\tilde \rho}  \wt g$.   On the other hand, if the charge and mass are free but the unperturbed system is exactly extremal, then the $\Delta \rho$ term drops out and the charge-to-mass ratio shift is $ \Delta z  = - \Delta g /\partial_{\tilde z}  \wt g $.  At the same time, the radial component of the metric $\wt g$ is by definition spacelike outside the horizon, so $\partial_{\tilde \rho}  \wt g >0$.   Moreover, since $\wt g$ dictates the gravitational potential at long distances, it decreases with $m$ and thus increases with the charge-to-mass ratio, so $\partial_{\tilde z} \wt g > 0$.  This logic implies that $\Delta \rho$ and $\Delta z$ have the same sign.  Since the entropy shift for a near-extremal black hole is dominated by the shift in the horizon, $\Delta S \sim \Delta \rho$, we discover that $\Delta S>0$, $\Delta \rho>0$, and $\Delta z>0$ are equivalent bounds.

Conveniently, the above logic immediately extends to the multi-charge generalization of the WGC proposed in \Ref{Cheung:2014vva}.  For a theory with multiple Abelian factors, the charge-to-mass ratio defines a vector $\mathbf{z}$ in charge space.  The WGC then mandates that the unit ball representing all possible large black holes be contained within the convex hull spanned by the set of all $ \mathbf z$ for the lighter states in the theory.  Crucially, for a multi-charged black hole, the perturbed metric only depends on the magnitude of its charge and not the direction.  Hence, \Eq{eq:gshift} still applies, provided we define $\wt z = |\mathbf{\wt z}|$ as the magnitude of the charge-to-mass ratio vector of the black hole and $\Delta z = \Delta \mathbf {z}  \cdot \mathbf{\wt z} / |\mathbf{\wt z}|$ as its shift.  Repeating exactly the argument of the previous paragraph, we learn that $\Delta S >0$, $\Delta \rho >0$, and $\Delta \mathbf {z}  \cdot \mathbf{\wt z}>0$ are all equivalent.  The last inequality implies that the extremality condition for a multi-charged extremal black hole is perturbed so that the corresponding unit ball is expanded outward in every direction, thus proving the multi-charge version of the WGC given by the convex hull condition in \Ref{Cheung:2014vva}.

\section{Discussion and  Conclusions} \label{sec:conclusions}

In this paper, we derived a positivity condition on classical corrections to the Bekenstein-Hawking entropy.
For near-extremal black holes this enforces positivity of a certain linear combination of coefficients of higher-dimension operators.
This very same combination of couplings corrects the extremality condition for black holes so that their charge-to-mass ratios approach unity from above for increasing size.  Large extremal black holes are thus unstable to decay to smaller extremal black holes.  Since the latter have charge-to-mass ratios greater than unity, they automatically fulfill the requirement of the WGC.

Our findings leave a number of avenues for future work. First, it would be interesting to determine if entropy considerations have any additional implications for the swampland program, for example by introducing new operators in extended theories like Einstein-dilaton gravity or by considering black holes embedded in asymptotically AdS or dS space, rotating black holes, or black holes of different topologies. Second, one would ideally like to understand the relationship, if any, between these entropy bounds and other contraints on low-energy dynamics coming from causality, analyticity, and unitarity.  Indeed, the positivity of entropy shifts discussed in this paper stems from state counting in the ultraviolet, which is highly reminiscent of dispersion relation bounds utilizing the positivity of forward cross-sections \cite{Adams:2006sv,Rattazzi,Cheung:2014ega,Bellazzini:2015cra,Cheung:2016yqr,Bellazzini:2016xrt,deRham:2017zjm,Bellazzini:2017fep} and amplitudes approaches based on the positivity of spectral representations \cite{GB+,Dvali:2012zc,Jenkins:2006ia}.

\vspace{1cm}

\begin{center}
 {\bf Acknowledgments}
\end{center}
We thank Nima Arkani-Hamed, Raphael Bousso, Sean Carroll, William Cottrell, Daniel Harlow, Tom Hartman, Yasunori Nomura, Hirosi Ooguri, Matthew Reece,  Prashant Saraswat, Gary Shiu, Leo Stein, and Aron Wall for helpful discussions.  C.C. and J.L. are supported by a Sloan Research Fellowship and DOE Early Career Award under grant no.~DE-SC0010255.   G.N.R. is supported by the Miller Institute for Basic Research in Science at the University of California, Berkeley.

\appendix

\section{Generalization to Arbitrary Dimension}\label{app:genD}

\subsection{Black Hole Spacetime}\label{sec:BHD}

In this appendix we generalize all of our results to arbitrary spacetime dimension $D\geq 4$. See footnote \ref{foot:massdim} for the mass dimensions of various quantities.  To begin, we consider the Reissner-Nordstr\"om metric in $D$ dimensions,
\be
{\rm d}s^2=-\wt f(r){\rm d}t^2 + \frac{1}{\wt g(r)}{\rm d}r^2 + r^2 {\rm d}\Omega_{D-2}^2,\label{eq:sphericalmetricD}
\ee
where ${\rm d}\Omega_{D-2}^2$ is line element on the unit $(D-2)$-sphere and
\be
\wt f(r) = \wt g(r) = 1-\frac{2\kappa^{2}M}{(D-2){{\Omega }_{D-2}}{{r}^{D-3}}}+\frac{{{Q}^{2}}\kappa^{2}}{(D-2)(D-3)\Omega _{D-2}^{2}{{r}^{2(D-3)}}},
\ee
as before denoting $\kappa^2 = 8\pi G$. The electromagnetic field strength is
\be
\wt{F}_{\mu\nu} {\rm d}x^\mu \wedge {\rm d}x^\nu = \frac{Q}{{{\Omega }_{D-2}}{{r}^{D-2}}}\mathrm{d}t\wedge \mathrm{d}r,
\ee
where the $(D-2)$-dimensional area of the unit codimension-two sphere is
\be
{{\Omega }_{D-2}}=\frac{2{{\pi }^{\frac{D-1}{2}}}}{\Gamma \left( \frac{D-1}{2} \right)}.
\ee
Next, let us define new variables for mass and charge in units of the Planck scale,
\eq{m = \frac{\kappa^2 M}{(D-2)\Omega_{D-2}}, \qquad 
q = \frac{\kappa Q}{\sqrt{(D-2)(D-3)}\Omega_{D-2}},
}{eq:unitsD}
along with a rescaled radial coordinate,
\eq{
x = r^{D-3},
}{}
in terms of which the metric component can be written simply as
\be
\wt g(r) = 1 -  \frac{2m}{x} + \frac{q^2}{x^2}.
\ee
The outer horizon is located at $x=\wt \chi = \wt \rho^{D-3}$, where
\be
\wt \chi = m + \sqrt{m^2 - q^2} = m(1+\xi)
\ee
and $\xi$ is defined as in \Eq{eq:xidef}. The extremality condition for the background Reissner-Nordstr\"om spacetime as before requires $q/m\leq 1$. The requirement of thermodynamic stability restricts our consideration to black holes with $\xi < \frac{D-3}{D-2}$, for which the specific heat is positive.

 Following the perturbative methods of Refs.~\cite{Kats:2006xp,Campanelli:1994sj}, we can compute the metric components at first order in perturbations, finding
\be
g(r)=1-\frac{2m}{x}+\frac{q^2}{x^2} -\frac{q^2}{x^{\frac{2(2D-5)}{D-3}}} \sum_{i=1}^8 \alpha_i c_i,
\label{eq:gD}
\ee
where the coefficients are
\be
\begin{aligned}
\alpha_1 &= \frac{(D-3)(D-4)}{D-2}\left[2\frac{13D^2-47D+40}{3D-7}q^2 - 8(3D-5)m x + 16(D-2) x^2 \right]\\
\alpha_2 &= 2\frac{D-3}{D-2}\Bigg[\frac{8D^3 - 55D^2 + 117D - 76}{3D-7}q^2 - 4(2D^2 - 10D + 11)m x \\&\qquad\qquad\qquad +2(3D-10)(D-2) x^2 \Bigg] \\
\alpha_3 &= 4\frac{D-3}{D-2}\Bigg[\frac{8D^3-48D^2+87D-44}{3D-7}q^2 -2(4D^2-17D+16)m x \\&\qquad\qquad\qquad + 8(D-2)(D-3)x^2 -2(D-2)(D-4)\frac{m^2 x^2}{q^2} \Bigg]\\
\alpha_4 &= 4(D-3)\left[\frac{(7D-13)(D-2)}{3D-7}q^2 - 2(3D-5)m x + 4(D-2) x^2 \right]\\
\alpha_5 &= 2(D-3)\left[\frac{(5D-9)(D-2)}{3D-7}q^2 - 2(2D-3)m x + 3(D-2) x^2\right]\\
\alpha_6 &= 4(D-3)\left[4\frac{(D-2)^2}{3D-7}q^2 - (3D-5)m x + 2(D-2) x^2 \right]\\
\alpha_7 &= 8\frac{(D-2)(D-3)^2}{3D-7}q^2\\
\alpha_8 &= 4\frac{(D-2)(D-3)^2}{3D-7}q^2 .
\end{aligned}\label{eq:alphas}
\ee

\subsection{Calculation of Entropy}

As before, the total entropy shift is $\Delta S = \Delta S_{\rm I} +\Delta S_{\rm H}$, where $\Delta S_{\rm I}$, defined in \Eq{eq:Sdyn}, arises from modifications of the low-energy graviton interactions and $\Delta S_{\rm H}$, defined in \Eq{eq:Sgeom}, is induced by the shift of the black hole horizon.  

To compute the entropy contribution from interactions, we substitute the unperturbed black hole background from \Sec{sec:BHD} into \Eq{eq:Waldpre}, yielding
\be
\begin{aligned}
\Delta S_{\rm I} &= \wt S \times \frac{2(D-3)}{m^{\frac{2}{D-3}} (1+\xi)^{\frac{D-1}{D-3}}}\Big\{4(D-2)\e_3\\&\qquad\;  -2(1-\xi)\left[(D-4)\e_1 + (D-3)\e_2 + 2(2D-5)\e_3+(D-2)\left(\e_4 + \frac{1}{2}\e_5 + \e_6\right)\right]\Big\}. \label{eq:WaldD}
\end{aligned}
\ee
To obtain the entropy contribution from the shift in the horizon, we apply \Eq{eq:delta_rho}.  The shift in the horizon area is then
\be
\Delta A = A - \wt A = (D-2)\Omega_{D-2}\wt \rho^{D-3}\Delta \rho = -\frac{(D-2)\Omega_{D-2} \wt \chi \Delta g(\wt \rho)}{\partial_{\tilde \rho} \wt g(\wt \rho) },
\ee
where the unperturbed area is $\wt A = \Omega_{D-2}\wt \rho^{D-2}$.
Inserting the perturbed metric in \Eqs{eq:gD}{eq:alphas}, we then obtain
\be
\begin{aligned}
\Delta S_{\rm H} =&\, \wt S \times  \frac{1}{(3D-7)m^{\frac{2}{D-3}}\xi(1+\xi)^{\frac{D-1}{D-3}}}\times \\&\hspace{20mm} \times \bigl\{  \e_1 (1-\xi)(D-3)(D-4)[(11D-24)\xi +D-4]  \\ &\hspace{22.4mm}+\e_2 (1-\xi)(D-3)[(10D^2 - 53D + 68)\xi + 2D^2 - 11D + 16] \\ &\hspace{22.4mm}+2\e_3 [-(16D^3 - 128 D^2 + 337D - 292)(1-\xi)^2 \\&\qquad\qquad\hspace{20mm} +2(3D-7)(4D^2 - 23 D + 32)(1-\xi) \\&\qquad\qquad\hspace{20mm} -2(D-2)(D-4)(3D-7)]\\&\hspace{22.4mm}+2\e_4(1-\xi)(D-2)(D-3)[5(D-2)\xi + D-4] \\ &\hspace{22.4mm}+2(\e_5+\e_6)(D-2)(D-3)(1-\xi)[2(D-2)\xi + D-3]  \\ &\hspace{22.4mm}  + 2(2\e_7+\e_8)(D-2)^2 (D-3)(1-\xi)^2 \bigr\}.
\end{aligned}\label{eq:dAD}
\ee
As before, we can consider a near-extremal limit in which $\xi\ll 1$ but $\Delta S \ll\wt S$ so that perturbation theory still applies.  This requires that
\be
\xi \gg \frac{|\e_i|}{m^{\frac{2}{D-3}}},\label{eq:ximinD}
\ee
which permits arbitrarily small $\xi$ for a sufficiently large black hole. However, for the classical higher-dimension operators to dominate, we again require $\rho \ll  1/\kappa  m_\phi^{D/2}$ according to \Eq{eq:rHbound}.  Hence, for a tree-level ultraviolet completion with $\e_i \sim 1/ m_\phi^2$, \Eq{eq:ximinD} becomes $\xi \gg \kappa^2  m_\phi^{D-2}$. 
Additionally, our argument in \Sec{sec:proof} imposes a further perturbativity criterion on the inverse temperature shift, $\Delta \beta \ll \wt \beta$. The background inverse temperature is
\be 
\wt \beta = \frac{2\pi m^{\frac{1}{D-3}}(1+\xi)^{\frac{D-2}{D-3}}}{(D-3)\xi},
\ee
while in the near-extremal limit the inverse temperature shift goes as $\Delta \beta\sim \e_i /\xi^3 m^{1/(D-3)}$. Hence, requiring $\Delta \beta \ll \wt \beta$, we have
\be
\xi \gg \frac{{|\e_i|^{1/2}}}{m^{\frac{1}{D-3}}}.
\ee
Along with the scaling of the $\e_i$ and the bound on $\rho$ imposed by \Eq{eq:rHbound}, this becomes just the requirement that
\be
\xi \gg  \kappa m_\phi^{(D-2)/2},
\ee
so $\xi$ can still be made parametrically small provided the heavy states are sub-Planckian. 

\subsection{New Positivity Bounds}

Combining \Eqs{eq:WaldD}{eq:dAD}, we obtain the total shift in entropy in $D$ dimensions,
\be
\begin{aligned}
\Delta S =&\,\wt S \times \frac{1}{(3D-7)m^{\frac{2}{D-3}}\xi(1+\xi)^{\frac{D-1}{D-3}}}\times \\ &\hspace{20mm} \times  \bigl\{\e_1 (D-3)(D-4)^2 (1-\xi)^2 \\ &\hspace{22.4mm}+\e_2 (D-3)(2D^2 - 11D+16) (1-\xi)^2 \\&\hspace{22.4mm} + 2\e_3 [(8D^3 - 60D^2 + 151D-128)(1-\xi)^2 \\&\qquad\qquad\hspace{20mm} -2(D-2)(2D-5)(3D-7)(1-\xi) \\&\qquad\qquad\hspace{20mm} +2(D-2)^2(3D-7)]\\&\hspace{22.4mm}+2\e_4(D-2)(D-3)(D-4)(1-\xi)^2 \\& \hspace{22.4mm} + 2\e_5 (D-2)(D-3)^2(1-\xi)^2 \\& \hspace{22.4mm}+ 2\e_6(D-2)(D-3)(1-\xi)[-2(2D-5)\xi + D - 3] \\&\hspace{22.4mm} + 4\e_7(D-2)^2(D-3)(1-\xi)^2 \\&\hspace{22.4mm} + 2\e_8 (D-2)^2 (D-3)(1-\xi)^2 \bigr\}.
\end{aligned}\label{eq:dSD}
\ee
Positivity of this entropy shift for all $\xi \in \left(0,\frac{D-3}{D-2}\right)$ then implies a family of new constraints on the higher-dimension operator coefficients, which generalizes \Eq{eq:bound4D},
\be
(1-\xi)^2 \e_0+ (D-2)^2 (3D-7) \xi \e_3 - \frac{1}{2}(D-2)(D-3)(3D-7)\xi(1-\xi)(2\e_3+\e_6) > 0, \label{eq:boundD}
\ee
where in analogy with \Eq{eq:d0def} we have defined
\be
\begin{aligned}
\e_0 &= \frac{1}{4}(D-3)(D-4)^2  \e_1 + \frac{1}{4}(D-3)(2D^2 - 11D+16)\e_2 \\ &\qquad + \frac{1}{2}(2D^3-16D^2+45D-44)\e_3 + \frac{1}{2}(D-2)(D-3)(D-4)\e_4 \\ & \qquad + \frac{1}{2}(D-2)(D-3)^2 (\e_5 + \e_6) + (D-2)^2 (D-3)\left(\e_7 + \frac{1}{2}\e_8\right).
\end{aligned}\label{eq:xyzD}
\ee
As before, the bound in \Eq{eq:boundD} is stronger than any finite set of bounds obtained for fixed values of $\xi$, i.e., each $\xi$ yields a linearly independent bound. As shown in \App{app:FD}, the bound in \Eq{eq:boundD} is field redefinition invariant for all values of $\xi$.

In the near-extremal limit, $\xi \ll1 $,  the bound in \Eq{eq:boundD} becomes
\be
\e_0>0. \label{eq:x}
\ee
The above inequality is closely related to the perturbation of the extremality condition discussed in \Sec{sec:WGC}.  Applying the same reasoning to general dimension $D$, we find that the extremality condition for the perturbed black hole is shifted by 
\be
\Delta z = \frac{4(D-3)}{(3D-7)(D-2)m^{\frac{2}{D-3}}}\e_0, \label{eq:extremal}
\ee
where $\e_0$ is exactly the same combination of coefficients defined in \Eq{eq:xyzD}.  Thus, the requirement of \Eq{eq:inequality} mandating positive entropy shift implies a constraint on the coefficients of higher-dimension operators that increases the charge-to-mass ratio of extremal black holes in the theory.  In turn, large black holes can decay to smaller black holes of a higher charge-to-mass-ratio, thus establishing the WGC in general dimension $D$.	

\section{Field Redefinition Invariance}\label{app:FD}

Any physical observable should be invariant under a reparameterization of the field variables.  Let us consider an arbitrary field redefinition,
\eq{
g_{\mu\nu} &\rightarrow  g_{\mu\nu} + \delta g_{\mu\nu},
}{}
where the perturbation is second order in derivatives, so 
\eq{
\delta g_{\mu\nu} &= r_1 R_{\mu\nu} + r_2 g_{\mu\nu}  R + r_3 \kappa^2  F_{\mu\rho}F_\nu^{\;\;\rho} +r_4 \kappa^2  g_{\mu\nu} F_{\rho\sigma}F^{\rho\sigma} 
}{eq:FD}
for a set of four arbitrary constants $r_i$.  Inserting this field redefinition into the action for Einstein-Maxwell theory induces new terms in the action proportional to the equations of motion,\footnote{The particular field redefinition in which the pure Einstein-Maxwell equations of motion are substituted into the higher-dimension operators is a special case of the transformation in \Eq{eq:FD}.} 
\be 
\delta {\cal L} = \frac{1}{2\kappa^2}\delta g^{\mu\nu} \left(R_{\mu\nu} - \frac{1}{2}R g_{\mu\nu}-\kappa^2 T_{\mu\nu}\right).
\ee
This has the net effect of shifting the higher-dimension operator coefficients in the action by
\be
\begin{aligned}
\e_1 &\rightarrow \e_1 -\frac{1}{4}r_1 - \frac{D-2}{4} r_2\\
\e_2 &\rightarrow \e_2 +\frac{1}{2}r_1\\
\e_3 &\rightarrow \e_3 \\
\e_4 &\rightarrow \e_4 + \frac{1}{8}r_1 + \frac{D-4}{8}r_2 -\frac{1}{4}r_3 -\frac{D-2}{4}r_4\\
\e_5 &\rightarrow \e_5 - \frac{1}{2}r_1 + \frac{1}{2}r_3 \\
\e_6 &\rightarrow \e_6 \\
\e_7 &\rightarrow \e_7 + \frac{1}{8}r_3 + \frac{D-4}{8}r_4\\
\e_8 &\rightarrow \e_8 - \frac{1}{2}r_3.
\end{aligned}\label{eq:citransform}
\ee
Because the field redefinition depends on four arbitrary constants, this reduces the naive basis of eight higher-dimension operator coefficients down to a set of four combinations that are automatically field redefinition invariant:
\be
\e_0, \e_3, \e_6, \e_9,
\ee
where $\e_0$ is defined in \Eq{eq:xyzD} and $\e_9 = \e_2 + \e_5 + \e_8$.    All physical quantities, like the bounds in \Eqs{eq:bound4D}{eq:boundD}, depend only on these combinations of coefficients.

 \noindent

\bibliographystyle{utphys-modified}
\bibliography{WGCproof}

\end{document}